\documentstyle [10pt,amsfonts] {article}
\input epsf

\newcommand{\beq}{\begin{equation}}
\newcommand{\eeq}{\end{equation}}
\newcommand{\beqs}{\begin{eqnarray}}
\newcommand{\eeqs}{\end{eqnarray}}

\catcode`@=11
\@addtoreset{equation}{section}
\@addtoreset{equation}{subsection}
\def\theequation{\ifnum\value{section}=0 \arabic{equation}\ignorespaces
\else \ifnum\value{section}=-1 A.\arabic{equation}\ignorespaces
\else \ifnum\value{subsection}=0 \thesection.\arabic{equation}\ignorespaces
\else \thesection.\arabic{subsection}.\arabic{equation}\ignorespaces
                           \fi
                      \fi
                 \fi}
\catcode`@=12

\begin{document}

\def\thefootnote{\fnsymbol{footnote}}

\baselineskip 6.0mm

\vspace{4mm}

\begin{center}

{\Large \bf Ground State Entropy of the Potts Antiferromagnet on Cyclic Strip 
Graphs}

\vspace{8mm}

\setcounter{footnote}{0}
Robert Shrock\footnote{email: robert.shrock@sunysb.edu}

\vspace{4mm}

Institute for Theoretical Physics \\
State University of New York       \\
Stony Brook, N. Y. 11794-3840  \\

\vspace{4mm}

\setcounter{footnote}{6}
Shan-Ho Tsai\footnote{email: tsai@hal.physast.uga.edu}

Department of Physics and Astronomy \\
University of Georgia \\
Athens, GA  30602 \\ 

\vspace{10mm}

{\bf Abstract}
\end{center}

We present exact calculations of the zero-temperature partition function
(chromatic polynomial) and the (exponent of the) ground-state entropy $S_0$ for
the $q$-state Potts antiferromagnet on families of cyclic and twisted cyclic
(M\"obius) strip graphs composed of $p$-sided polygons.  Our results suggest a
general rule concerning the maximal region in the complex $q$ plane to which
one can analytically continue from the physical interval where $S_0 > 0$.  The
chromatic zeros and their accumulation set ${\cal B}$ exhibit the rather
unusual property of including support for $Re(q) < 0$ and provide further
evidence for a relevant conjecture.

\vspace{16mm}

\pagestyle{empty}
\newpage

\pagestyle{plain}
\pagenumbering{arabic}
\renewcommand{\thefootnote}{\arabic{footnote}}
\setcounter{footnote}{0}

The $q$-state Potts antiferromagnet (AF) \cite{potts,wurev} exhibits nonzero
ground state entropy, $S_0 > 0$ (without frustration) for sufficiently
large $q$ on a given graph or lattice.  This is equivalent to a ground state
degeneracy per site $W > 1$, since $S_0 = k_B \ln W$.  Such nonzero ground
state entropy is important as an exception to the third law of
thermodynamics \cite{al}.  There is a close connection with graph theory here,
since the zero-temperature partition function of the above-mentioned $q$-state
Potts antiferromagnet on a graph $G$ satisfies $Z(G,q,T=0)_{PAF}=P(G,q)$, where
$P(G,q)$ is the chromatic polynomial expressing the number of ways of coloring
the vertices of the graph $G$ with $q$ colors such that no two adjacent
vertices have the same color \cite{bl}-\cite{rtrev}.  Thus
\beq 
W(\{G\},q) = \lim_{n \to \infty} P(G,q)^{1/n}
\label{w}
\eeq 
where $n=v(G)$ is the number of vertices of $G$ and $\{G\} = \lim_{n \to
\infty}G$. \footnote{Some previous works include Refs. \cite{bl}-\cite{bax}.}
Since $P(G,q)$ is a polynomial, one can generalize $q$ from ${\mathbb Z}_+$ to
${\mathbb R}$ and to ${\mathbb C}$, and this is useful, just as the study of
functions of a complex variable gives deeper insight into functions of a real
variable in mathematics.  The zeros of $P(G,q)$ in the complex $q$ plane,
called chromatic zeros, are of basic importance. Their accumulation set in the
limit $n \to \infty$, denoted ${\cal B}$, is the continuous locus of points
where $W(\{G\},q)$ is nonanalytic.  A fundamental question concerning the Potts
antiferromagnet is the maximal region in the complex $q$ plane to which one can
analytically continue the function $W(\{G\},q)$ from physical values where
there is nonzero ground state entropy, i.e., $W > 1$.  We denote this region as
$R_1$.  In the present work we present exact calculations of $P(G,q)$ and
$W(\{G\},q)$ for families of strip graphs with free transverse and periodic 
longitudinal boundary conditions.  From these we infer an answer to the above 
question.

These results are of further interest because of a property of the chromatic
zeros.  For many years, no examples of chromatic zeros were found with negative
real parts, leading to the conjecture that $Re(q) \ge 0$ for any chromatic zero
\cite{farrell}.  Although this was later shown to be false \cite{read91}, very
few cases of graphs with chromatic zeros having $Re(q) < 0$ are known, and the
investigation of such cases is thus valuable for the insight it yields into
properties of chromatic zeros.  Note that the condition that a graph has some
chromatic zeros with $Re(q) < 0$ is a necessary but not sufficient condition
that it has an accumulation set ${\cal B}$ with support for $Re(q) < 0$.  For
the graph families considered here we find that both the chromatic zeros and
their accumulation set ${\cal B}$ include support for $Re(q) < 0$.

We start with a cyclic strip of the square lattice comprised of $m$ squares,
with its longitudinal (transverse) direction taken to be horizontal (vertical).
By cyclic, we mean that this strip has periodic boundary conditions in the
longitudinal direction, which will extend to infinity in the $m \to \infty$
limit.  Now add $k-2$ vertices to the upper edge and $k-2$ vertices to the
lower edge of each square.  We denote this graph as as $(Ch)_{k,m,cyc.}$.  The
analogous graph with twisted periodic longitudinal boundary conditions is a
M\"obius strip, denoted $(Ch)_{k,m,cyc.,t}$.  These graphs have $n=2(k-1)m$
vertices and can be regarded as cyclic and twisted cyclic strips of $m$
$p$-sided polygons, where $p=2k$, such that each $p$-gon intersects the
previous one on one of its edges, and intersects the next one on its opposite
edge.  For a given $m$, the $(Ch)_{k,m,cyc.}$ and $(Ch)_{k,m,cyc.,t}$ form
separate homeomorphic families.\footnote{Two graphs $G$ and $H$ are
homeomorphic to each other if one of them, say $H$, can be obtained from the
other, $G$, by successive insertions of degree-2 vertices on bonds of $G$ 
(e.g., \cite{rw,wa3,hs}).} The girth (length of minimum closed path) is
$g=p=2k$ for both.

Define
\beq
D_k = \sum_{s=0}^{k-2}(-1)^s {{k-1}\choose {s}} q^{k-2-s}
\label{dk}
\eeq
By iterated use of the deletion-contraction theorem 
\cite{rrev}, we obtain the chromatic polynomials
\beq
P((Ch)_{k,m,cyc.},q) = c_0 + (c_1)^m +(q-1)[ (c_2)^m + (c_3)^m ]
\label{pch}
\eeq
\beq
P((Ch)_{k,m,cyc.,t},q) = c_0^{(t)} + (c_1)^m + 
(-1)^k(q-1)[ (c_2)^m - (c_3)^m ]
\label{pchtw}
\eeq
with $c_0=q^2-3q+1$, $c^{(t)}_0=-1$, 
\beq
c_1=D_{2k}
\label{c1}
\eeq
\beq
c_2=(-1)^{k+1}D_{k+1}+D_k
\label{c2}
\eeq
\beq
c_3=(-1)^{k+1}D_{k+1}-D_k
\label{c3}
\eeq
In the lowest case, $k=2$ (cyclic and twisted cyclic ladder graphs),
eqs. (\ref{pch}), (\ref{pchtw}) reduce to known results \cite{bds,read91}.  In
Ref. \cite{w} we have calculated ${\cal B}$ and $W$ for these cases, as part of
the general studies in Refs. \cite{wa3}-\cite{wa2}.  $P$ always has a factor
$q(q-1)$; in addition, $P((Ch)_{k,m,cyc.},q)$ has a $(q-2)$ factor for
$(k,m)=(e,o)$ and $P((Ch)_{k,m,cyc.,t},q)$ has a factor $D_k^2$ for odd $k$,
$(q-2)D_k^2$ for $(k,m)=(e,e)$ and $D_k^2$ for $(e,o)$, where $e=$ even and
$o=$ odd. (For odd $k$, $D_k$ has a factor $(q-2)$.)  These are in accord with
the values of the chromatic number $\chi$ (minimum $q$ to color the graph with
the above constraint): for $(k,m)=(e,e)$, $(o,e)$, and $(o,o)$, 
$\chi \Bigl ((Ch)_{k,m,cyc.}\Bigr )=2$ and 
$\chi \Bigl ((Ch)_{k,m,cyc.,t}\Bigr) =3$ while for $(k,m)=(e,o)$, 
$\chi \Bigl ((Ch)_{k,m,cyc.}\Bigr )=3$ and 
$\chi \Bigl ((Ch)_{k,m,cyc.,t}\Bigr )=2$. 

Taking $m$ and hence $n$ to infinity, we determine $W$.  For special points
$q_s$ (e.g., $q_s=0,1$) where the limits $n \to \infty$ and $q \to q_s$ do not
commute, we use the order of limits in eq.  (1.5) of Ref. \cite{w}, i.e. $n \to
\infty$, then $q \to q_s$.  For a given $q \in {\mathbb C}$, $W(q)$ is
determined by the term $c_j$ which is ``leading'', i.e., has maximal $|c_j| >
1$ over the $j$'s, and if $|c_j| < 1$, then $W$ is determined by $c_0$, and
$|W|=1$.  The locus ${\cal B}$ is determined by the degeneracy of leading
$c_j$'s.  From eqs. (\ref{pch}), (\ref{pchtw}), it follows that $W$ and ${\cal
B}$ are the same for $(Ch)_{k,m=\infty,cyc.}$ and $(Ch)_{k,m=\infty,cyc.,t}$
(indicated in figures by $(t)$). 

\begin{figure}
\vspace{-4cm}
\centering
\leavevmode
\epsfxsize=3.0in
\begin{center}
\leavevmode
\epsffile{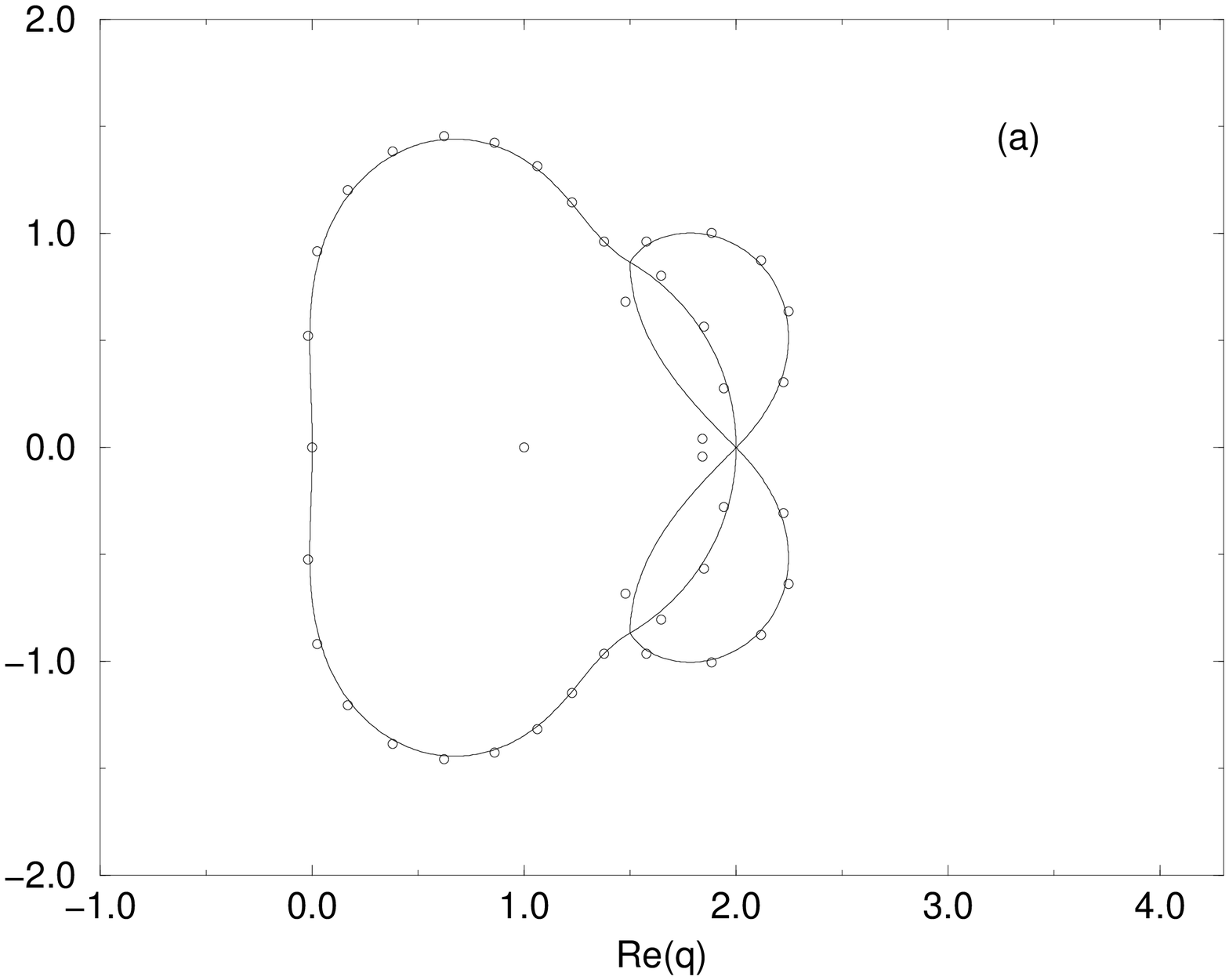}
\end{center}
\vspace{-2cm}
\begin{center}
\leavevmode
\epsfxsize=3.0in
\epsffile{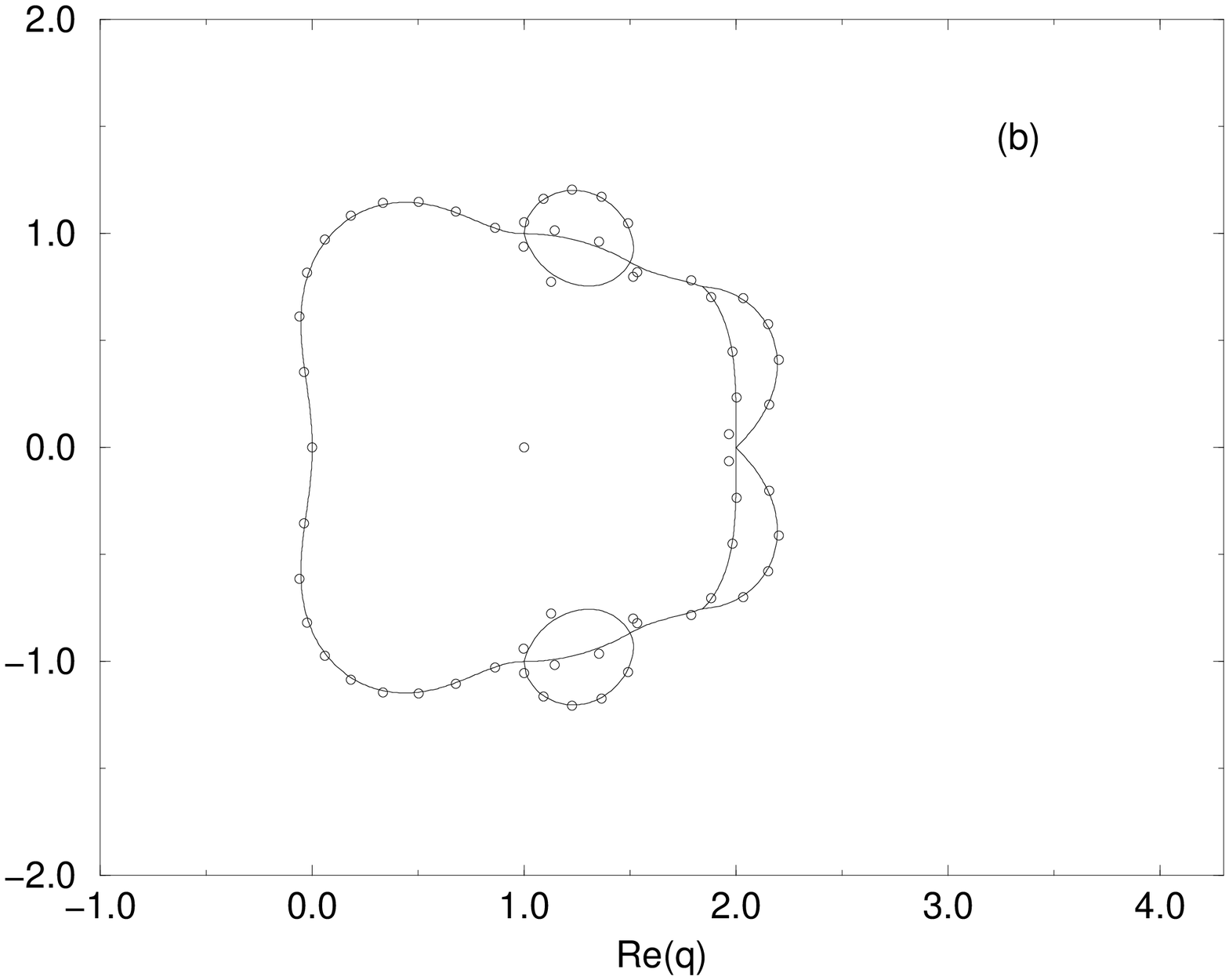}
\end{center}
\vspace{-2cm}
\caption{\footnotesize{${\cal B}$ for $\lim_{m \to \infty}(Ch)_{k,m,cyc.(t)}$
with $k=$ (a) 3 (b) 4. Chromatic zeros are shown for the cyclic case with 
$m=10$, i.e., $n=$ (a) 40 (b) 60.}}
\end{figure}

\pagebreak

\begin{figure}
\vspace{-4cm}
\centering
\leavevmode
\epsfxsize=3.0in
\begin{center}
\leavevmode
\epsffile{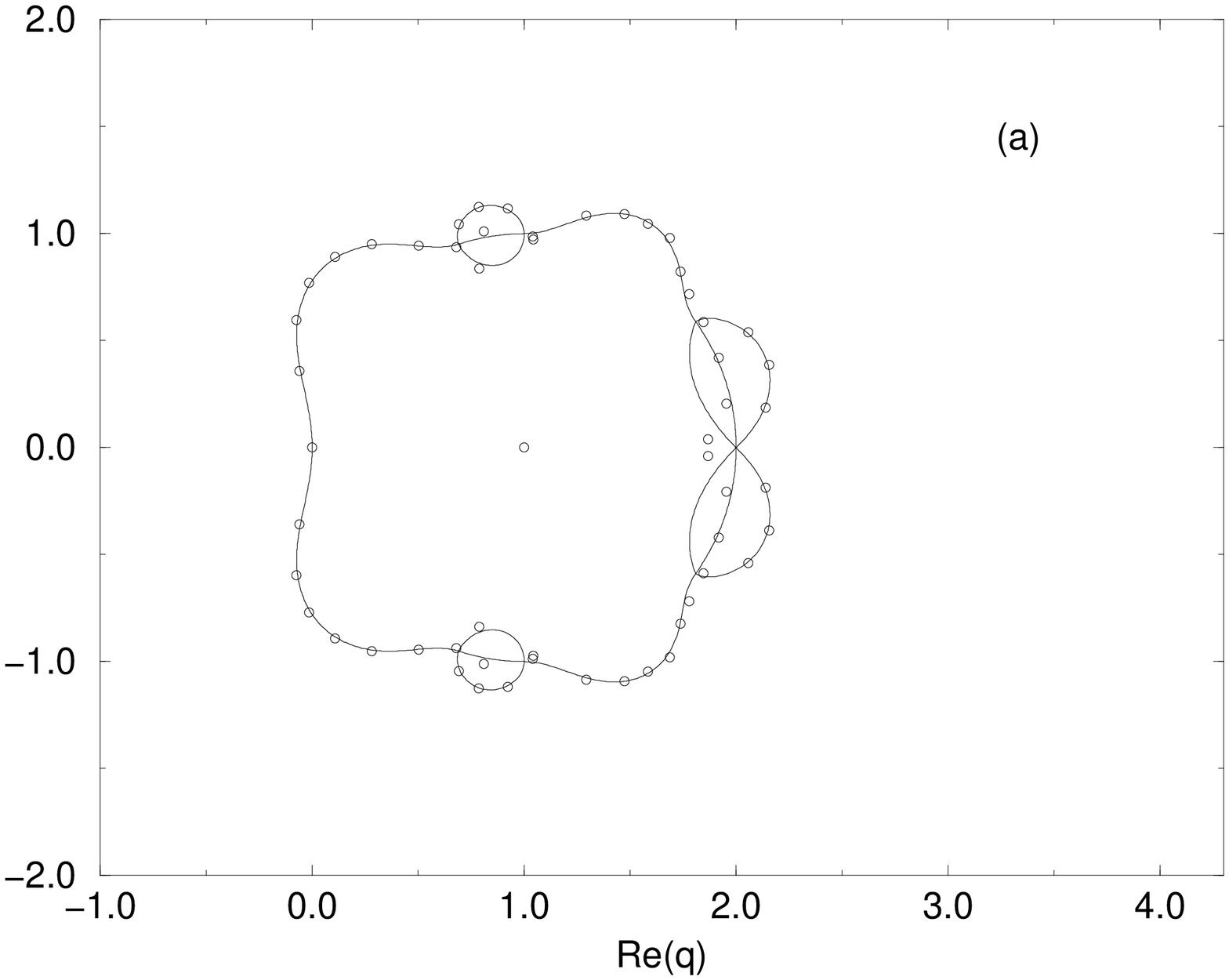}
\end{center}
\vspace{-2cm}
\begin{center}
\leavevmode
\epsfxsize=3.0in
\epsffile{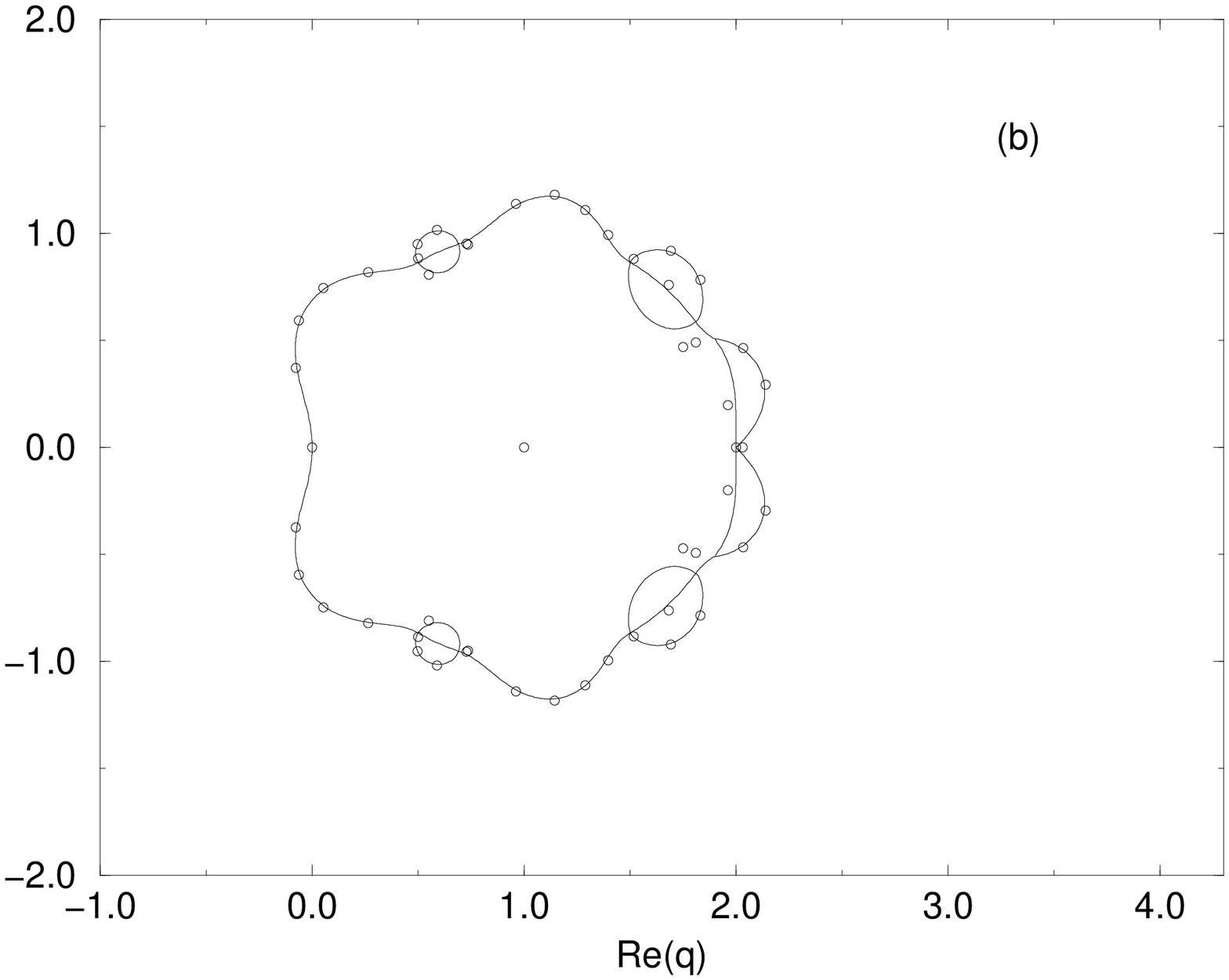}
\end{center}
\vspace{-2cm}
\caption{\footnotesize{${\cal B}$ for $\lim_{m \to \infty}(Ch)_{k,m,cyc.(t)}$ 
with $k=$ (a) 5 (b) 6. Chromatic zeros for are shown for the cyclic case with 
(a) $m=8$ ($n=64$), (b) $m=5$ ($n=50$).}}
\end{figure}

This locus is shown in Figs. 1 and 2 for $3 \le k \le 6$ together with
chromatic zeros for long finite strips for comparison.  We note the following
theorems: ${\cal B}$ (i) separates the $q$ plane into various regions; (ii) is
compact in this plane; and (iii) for $k \ge 3$, contains support for $Re(q) <
0$.  Thm. (i) is proved by explicit solution of the equations for the
boundary ${\cal B}$.  Thm. (ii) is proved by recasting the degeneracy
equations in the variable $z=1/q$ and showing that they have no solution for
$z=0$.  Thm. (iii) is proved below.

We find the following regions and forms for $W$: first, $R_1$, which includes 
the real intervals $q \ge 2$  and $q < 0$ and surrounds all of ${\cal B}$. Here
\beq
W = (D_{2k})^{\frac{1}{2(k-1)}} \quad {\rm for} \quad q \in R_1
\label{wr1}
\eeq
For real $q \ge 2$ and a given $k$, $W$ increases monotonically from 1 at 
$q=2$, approaching $q$ from below as $q \to \infty$: 
$W(q \to \infty) = q[1 - (2k-1)(2k-2)^{-1}q^{-1} + O(q^{-2})]$.  For a fixed 
$q$ in this interval, $W$ is a monotonically increasing function of $k$.
This can be understood physically in the case of integral $q \ge 2$ since an
increase in $k$ increases the girth and thereby weakens the coloring
constraint.  As $q \to 0^-$ in $R_1$, $W \to 
(2k-1)^{\frac{1}{2(k-1)}}$.  For regions $R_j \ne R_1$, only the magnitude 
$|W|$ can be determined \cite{w}. The innermost region, denoted $R_2$, 
includes the interval $0 < q < 2$.  Here, respectively, 
\beq
|W| = |c_{2,3}|^{\frac{1}{2(k-1)}}  \quad {\rm for} \quad q \in R_2 \quad 
{\rm and} \quad k \quad {\rm even, \ odd} 
\label{wr2}
\eeq
The condition of degeneracy of leading terms in the vicinity of $q=0$ is 
$|c_1|=|c_2|$ for even $k$ and $|c_1|=|c_3|$ for odd $k$.  Expanding these for
fixed $k$ and $r \to 0$, where $q=re^{i\theta}$, we get 
\beq
\cos \theta = -\frac{(k^2-3k+1)r}{2(2k-1)} + O(r^2) 
\label{costheta}
\eeq
Hence, in the vicinity of $q=0$, the curve ${\cal B}$ is concave toward the
right (bends to the upper and lower right) for $k=2$ but is concave to the left
for $k \ge 3$. This proves Thm. (iii), which, in turn, implies that for a given
$k \ge 3$, and for sufficiently large $n$, these families have chromatic zeros
with $Re(q) < 0$ (as is evident from Figs. 1 and 2), which become dense, as $n
\to \infty$, to form the part of the respective ${\cal B}$ with $Re(q) < 0$.
For $k=3,4$ and 5, chromatic zeros of $(Ch)_{k,m,cyc.}$ and
$(Ch)_{k,m,cyc.,t}$ with $Re(q) < 0$ occur first for strip lengths $m=6,4$,
and 3, respectively; when $k$ reaches the value $k=6$, they occur already for
the minimum nontrivial strip length, $m=2$.

The fact that these families combine the properties (ii) and
(iii) is of particular interest since we have previously found a number of
families of graphs with $Re(q) < 0$ for some chromatic zeros and part of 
${\cal B}$, but in these cases, ${\cal B}$ was noncompact (unbounded in the $q$
plane) \cite{wa2}.  The property (i) also contrasts with the situation for
homogeneous open strips, where we found \cite{strip,hs} that ${\cal B}$ does
not enclose regions in the $q$ plane. 

We next comment on the other regions.  At the point $q=2$ ($\equiv q_c$, where
$q_c$ is the maximal value of real $q \in {\cal B}$ \cite{p3afhc,w}) one branch
of ${\cal B}$ crosses the real axis vertically. To the upper and lower right of
$q_c$, there are two complex-conjugate (c.c.) regions, denoted $R_{q_c,r}$ and
$R_{q_c,r}^*$ ($r=$ ``right''); here, $|W|=|c_3|^{1/[2(k-1)]}$ for even $k$ and
$|W|=|c_2|^{1/[2(k-1)]}$ for odd $k$.  In the latter case of odd $k$, there are
two additional c.c. regions adjacent to $q_c$ to the upper and lower left,
denoted $R_{q_c; \ell}$, $R_{q_c; \ell}^*$ ($\ell=$ ``left''); in these
regions, $|W|=1$.  Thus, for even and odd $k$, four and six curves on ${\cal
B}$ intersect at $q_c$, respectively.  As is evident from Figs. 1 and 2, for $k
\ge 4$ there are further c.c. pairs of regions, each consisting of a pair on
either side of the ``main'' part of ${\cal B}$; here, in the outer parts,
$|W|=|c_3|^{1/[2(k-1)]}$ for even $k$ and $|W|=|c_2|^{1/[2(k-1)]}$ for odd $k$,
while in the inner parts, $|W|=1$ for both even and odd $k$.  From our findings
here we infer that the total number of regions is $2k$.

Define the outer envelope ${\cal E}$ of ${\cal B}$ to be the set of $q \in
{\cal B}$ with maximal value of $|q-1|$; this is the inner boundary of
region $R_1$.  We observe that this envelope ${\cal E}$ always lies outside of
the unit circle $|q-1|=1$.  As $k$ increases, ${\cal B}$ lies closer to
this circle.  

Our present results strengthen the evidence for our conjectures
\cite{strip,wa2} that on a graph with well-defined lattice directions, a
necessary property for there to be chromatic zeros and, in the $n \to \infty$
limit, a locus ${\cal B}$ including support for $Re(q) < 0$,
is that the graph has at least one global circuit, defined as a route along a
lattice which is topologically equivalent to the circle, $S^1$.  (This is known
not to be a sufficient property, as shown, e.g., by the circuit and ladder
graphs, which have such global circuits, but whose chromatic zeros and loci
${\cal B}$ have support only for $Re(q) \ge 0$.)  Note that in the second 
conjecture, the length of this global circuit, $L_{g.c.}$,
must $\to \infty$ as $n \to \infty$ in order for some chromatic zeros and part
of the locus ${\cal B}$ to include support for $Re(q) < 0$.  For the family
$(Ch)_{k,m,cyc.}$ there are two such global circuits, along the upper and lower
sides of the strip.\footnote{We have also calculated $P$, $W$, and ${\cal B}$
for the cyclic strips of the square lattice with width $L_y=3$ and the kagom\'e
lattice with $L_y=2$, and again these yield chromatic zeros and ${\cal B}$ with
support for $Re(q) < 0$ and have ${\cal B}$ separating the $q$ plane into
different regions.}

Our present study, combined with our earlier calculations \cite{w}-\cite{wa2},
suggests an answer to the basic question of how large is the region $R_1$ to
which one can analytically continue $W$ from the interval in $q$ where $S_0 >
0$ for graphs with regular lattice directions: a sufficient condition that in
the $n\to \infty$ limit the locus ${\cal B}$ separates the $q$ plane into two
or more regions is that the graph has a global circuit with $\lim_{n \to
\infty} \ell_{g.c.} = \infty$.\footnote{That this is is not a necessary
condition is shown by our results on inhomogeneous open strip graphs
\cite{strip}.  Moreover, ${\cal B}$ includes the point $q=0$ for the families
$(Ch)_{k,m=\infty,cyc.}$; however, our results in Ref. \cite{wc,wa,wa2}
show, and explain why, in cases where there are global circuits, ${\cal B}$
does not necessarily include the point $q=0$.} Thus, for graphs (with regular
lattice directions), a necessary condition that $R_1$ includes the full $q$ 
plane (except for the set of measure zero occupied by ${\cal B}$) is that the 
graphs do not contain any such global circuits.

This research was supported in part by the NSF grant PHY-97-22101.

\vfill
\eject

\vspace{6mm}

\begin{thebibliography}{99}

\bibitem{potts}{Potts, R. B. 1952 Proc. Camb. Phil. Soc. {\bf 48}, 106.}

\bibitem{wurev}{Wu, F. Y. 1982 Rev. Mod. Phys. {\bf 54}, 235; 
Wu, F. Y. 1983 {\it ibid}. {\bf 55}, 315 (errata).}

\bibitem{al}{Aizenman, M. and Lieb, E. H. 1981 J. Stat. Phys. {\bf 24}, 279; 
Chow, Y. and Wu, F. Y. 1987 Phys. Rev. {\bf B36}, 285.}

\bibitem{bl}{Birkhoff, G. D. and Lewis, D. C. 1946 Trans. Am. Math. Soc. 
{\bf 60}, 355.}

\bibitem{rrev}{Read, R. C. 1968 J. Combin. Theory {\bf 4}, 52.}

\bibitem{rtrev}{Read, R. C. and Tutte, W. T. 1988 ``Chromatic Polynomials'',
in {\it Selected Topics in Graph Theory, 3}, (Academic Press, NY).}

\bibitem{lieb}{E. H. Lieb, Phys. Rev. {\bf 162}, 162 (1967).}

\bibitem{bds}{Biggs, N. L., Damerell, R. M. and Sands, D. A. 1972 
J. Combin. Theory B {\bf 12}, 123 (1972).}

\bibitem{bkw}{Beraha, S., Kahane, J., and Weiss, N. 1980 J. Combin. Theory B
{\bf 28}, 52.}

\bibitem{bax}{Baxter, R. J. 1987 J. Phys. A {\bf 20}, 5241.}

\bibitem{farrell}{Farrell, E. J. 1980 Discrete Math. {\bf 29}, 161.}

\bibitem{read91}{Read, R. C. and Royle, G. F. 1991 in {\it Graph Theory,
Combinatorics, and Applications} (Wiley, NY), vol. 2, p. 1009.}

\bibitem{rw}{Read, R. C. and Whitehead, E. G., Discrete Math. in press.} 

\bibitem{wa3}{Shrock, R. and Tsai, S.-H. 1998 J. Phys. A {\bf 31}, 9641.}

\bibitem{hs}{Shrock, R. and Tsai, S.-H. 1998 Physica {\bf A259}, 315.}

\bibitem{p3afhc}{Shrock, R. and Tsai, S.-H. 1997 J. Phys. A {\bf 30}, 495.}

\bibitem{w}{Shrock, R. and Tsai, S.-H. 1997 Phys. Rev. {\bf E55}, 5165.}

\bibitem{wc}{Shrock, R. and Tsai, S.-H. 1997 Phys. Rev. {\bf E56}, 1342, 
4111.}

\bibitem{wa}{Shrock, R. and Tsai, S.-H. 1997 Phys. Rev. {\bf E56}, 3935.}

\bibitem{w2d}{Shrock, R. and Tsai S.-H. 1998 Phys. Rev. {\bf E58}, 4332.}

\bibitem{strip}{Ro\v{c}ek, M., Shrock, R., and Tsai, S.-H. 1998 Physica 
{\bf A252}, 505; {\it ibid.} {\bf A259}, 367.}

\bibitem{wa2}{Shrock, R. and Tsai, S.-H. 1999 Physica {\bf A265}, 186.} 

\end{thebibliography}
\end{document}